\documentclass[twocolumn]{aastex63}

\usepackage{lipsum}
\usepackage{amsmath}
\usepackage{amssymb}
\usepackage{empheq}
\usepackage{mathrsfs}
\usepackage{textcomp}
\usepackage{graphicx}
\usepackage{hyperref}
\usepackage{graphicx}
\usepackage{xspace}
\usepackage{enumitem}
\usepackage[caption=false]{subfig}
\usepackage{multirow}
\usepackage{longtable}
\hypersetup{colorlinks, linkcolor={blue}, citecolor={blue}, urlcolor={blue}}

\usepackage{lineno}

\def\bea{\begin{eqnarray}}
\def\eea{\end{eqnarray}}

\definecolor{MyB}{rgb}{0.1,0.1,1.0}
\definecolor{MyB2}{rgb}{0.0,0.0,0.7}

\newcommand*{\rom}[1]{\expandafter\@slowromancap\romannumeral #1@}
\makeatother

\begin{document}
\title{A Uniform Type Ia Supernova Distance Ladder with the Zwicky Transient Facility: \\Absolute Calibration Based on the Tip of the Red Giant Branch (TRGB) Method}
\author[0000-0002-2376-6979]{Suhail Dhawan}
\email{suhail.dhawan@ast.cam.ac.uk}
\affiliation{Institute of Astronomy and Kavli Institute for Cosmology, University of Cambridge, Madingley Road, Cambridge CB3 0HA, UK}
\author[0000-0002-4163-4996]{Ariel Goobar}
\affiliation{The Oskar Klein Centre for Cosmoparticle Physics, Department of Physics, Stockholm University, SE-10691 Stockholm, Sweden}
\author[0000-0001-5975-290X]{Joel Johansson}
\affiliation{The Oskar Klein Centre for Cosmoparticle Physics, Department of Physics, Stockholm University, SE-10691 Stockholm, Sweden}
\author{In Sung Jang}
\affiliation{Department of Astronomy \& Astrophysics \& Kavli Institute for Cosmological Physics, University of Chicago, 5640 South Ellis Avenue, Chicago, IL 60637, USA}
\author{Mickael Rigault}
\affiliation{Univ Lyon, Univ Claude Bernard Lyon 1, CNRS/IN2P3, IP2I Lyon, UMR 5822, F-69622, Villeurbanne, France}
\author{Luke Harvey}
\affiliation{School of Physics, Trinity College Dublin, The University of Dublin, Dublin 2, Ireland}
\author{Kate Maguire}
\affiliation{School of Physics, Trinity College Dublin, The University of Dublin, Dublin 2, Ireland}

\author[0000-0003-3431-9135]{Wendy~L.~Freedman}\affil{Department of Astronomy \& Astrophysics \& Kavli Institute for Cosmological Physics, University of Chicago, 5640 South Ellis Avenue, Chicago, IL 60637, USA}
\author[0000-0002-1576-1676]{Barry~F.~Madore}
\affil{The Observatories of the Carnegie Institution for Science, 813 Santa Barbara St., Pasadena, CA 91101, USA}
\author[0000-0002-3321-1432]{Mathew Smith}
\affiliation{Univ Lyon, Univ Claude Bernard Lyon 1, CNRS/IN2P3, IP2I Lyon, UMR 5822, F-69622, Villeurbanne, France}
\author[0000-0003-1546-6615]{Jesper Sollerman}
\affiliation{The Oskar Klein Centre, Department of Astronomy, Stockholm University, SE-10691 Stockholm, Sweden}
\author[0000-0002-1031-0796]{Young-Lo Kim}
\affiliation{Department of Physics, Lancaster University, Lancs LA1 4YB, UK}
\author{Igor Andreoni }
\affiliation{Department of Astronomy, University of Maryland, College Park, MD 20742, USA}
\author[0000-0001-8018-5348]{Eric C. Bellm}
\affiliation{DIRAC Institute, Department of Astronomy, University of Washington, 3910 15th Avenue NE, Seattle, WA 98195, USA}
\author[0000-0002-8262-2924]{Michael W. Coughlin}
\affiliation{School of Physics and Astronomy, University of Minnesota,
Minneapolis, Minnesota 55455, USA}
\author[0000-0002-5884-7867]{Richard Dekany}
\affiliation{Caltech Optical Observatories, California Institute of Technology, Pasadena, CA 91125, USA}

\author{Matthew J. Graham}
\affiliation{Division of Physics, Mathematics, and Astronomy, California Institute of Technology, Pasadena,USA}
\author{Shrinivas R. Kulkarni}
\affiliation{Division of Physics, Mathematics, and Astronomy, California Institute of Technology, Pasadena,USA}
\author[0000-0003-2451-5482]{Russ R. Laher}
\affiliation{IPAC, California Institute of Technology, 1200 E. California
             Blvd, Pasadena, CA 91125, USA}
\author{Michael S. Medford}
\affiliation{University of California, Berkeley, Department of Astronomy, Berkeley, CA 94720}
\affiliation{Lawrence Berkeley National Laboratory, 1 Cyclotron Rd., Berkeley, CA 94720}
\author{James D. Neill}
\affiliation{Division of Physics, Mathematics, and Astronomy, California Institute of Technology, Pasadena,USA}
\author{Guy Nir}
\affiliation{Department of Astronomy, University of California, Berkeley, CA 94720-3411, USA}
\author{Reed Riddle}
\affiliation{Caltech Optical Observatories, California Institute of Technology, Pasadena, CA 91125, USA}
\author[0000-0001-7648-4142]{Ben Rusholme}
\affiliation{IPAC, California Institute of Technology, 1200 E. California
             Blvd, Pasadena, CA 91125, USA}

\begin{abstract}
The current Cepheid-calibrated distance ladder measurement of $H_0$ is reported to be in tension with the values inferred from the cosmic microwave background (CMB), assuming standard cosmology. However,  some tip of the red giant branch (TRGB)  estimates report $H_0$ in better agreement with the CMB. Hence, it is critical to reduce systematic uncertainties in local measurements to understand the Hubble tension. In this paper, we propose a uniform distance ladder  between the second and third rungs, combining SNe~Ia observed by the Zwicky Transient Facility (ZTF) with a TRGB calibration of their absolute luminosity. A large, volume-limited sample of both calibrator and Hubble flow SNe~Ia from the \emph{same} survey minimizes two of the largest sources of systematics: host-galaxy bias and non-uniform photometric calibration.  We present results from a pilot study using existing TRGB distance to the host galaxy of ZTF SN~Ia SN 2021rhu (aka ZTF21abiuvdk)  in NGC7814. Combining the ZTF calibrator with a volume-limited sample from the first data release of ZTF Hubble flow SNe~Ia, we infer $H_0 = 76.94 \pm 6.4\, {\rm km}\,{\rm s^{-1}}\,{\rm Mpc^{-1}}$, an $8.3 \%$ measurement. The error budget is dominated by the single object calibrating the SN~Ia luminosity in this pilot study. However, the ZTF sample includes already five other SNe~Ia  within $\sim$ 20 Mpc for which TRGB distances can be obtained with HST. Finally, we present the prospects of building this distance ladder out to 80 Mpc with JWST observations of more than one hundred ZTF SNe~Ia.
\end{abstract}

\section{Introduction} 
In recent years, a remarkable increase in
accuracy obtained by a broad range of independent cosmological observations has provided compelling support for our current standard $\Lambda$ cold dark
matter ($\Lambda$CDM) model. This concordance cosmology successfully explains the measurements of fluctuations in the temperature and polarization of the cosmic microwave background (CMB) radiation \citep{2020A&A...641A...6P} as well as observations of large-scale structure and matter fluctuations
in the universe, e.g. baryon acoustic oscillations \citep[BAO;][]{2019MNRAS.486.2184M}.

With improved accuracy of recent observations, some discrepancies have been noted. The {\it prima facie} most significant tension is between the CMB inferred value of the Hubble constant ($H_0$) and the direct measurement of its local value \citep{riess2021:h0}. The local measurements are based on a calibration of the absolute luminosity of Type Ia supernovae (SNe~Ia) using independent distances to host galaxies of nearby SNe~Ia, known as the ``cosmic distance ladder".
This claimed tension, if confirmed, it could provide evidence for of new fundamental physics beyond the standard model of cosmology. It could, however, be a sign of unknown sources of systematic error. 
Currently, the local $H_0$ methods have slight differences in their values. The tip of the red giant branch \citep[TRGB;][]{2021ApJ...919...16F} and Cepheid \citep{riess2021:h0} distance scales yield values of 69.8 $\pm$ 1.7  \citep[however, see also,][for, e.g.]{Blakeslee2021} and 73.04 $\pm$ 1.04 km${\rm s}^{-1}{\rm Mpc}^{-1}$, respectively. 
Understanding these differences is important to discern whether the tension is a sign of novel physics or a yet-to-be-revealed systematic error. To date, only the TRGB and Cepheid measurements have measured distances to ten or more host galaxies of SNe~Ia.

\begin{figure*}
    \centering

    \includegraphics[width=.95\textwidth]{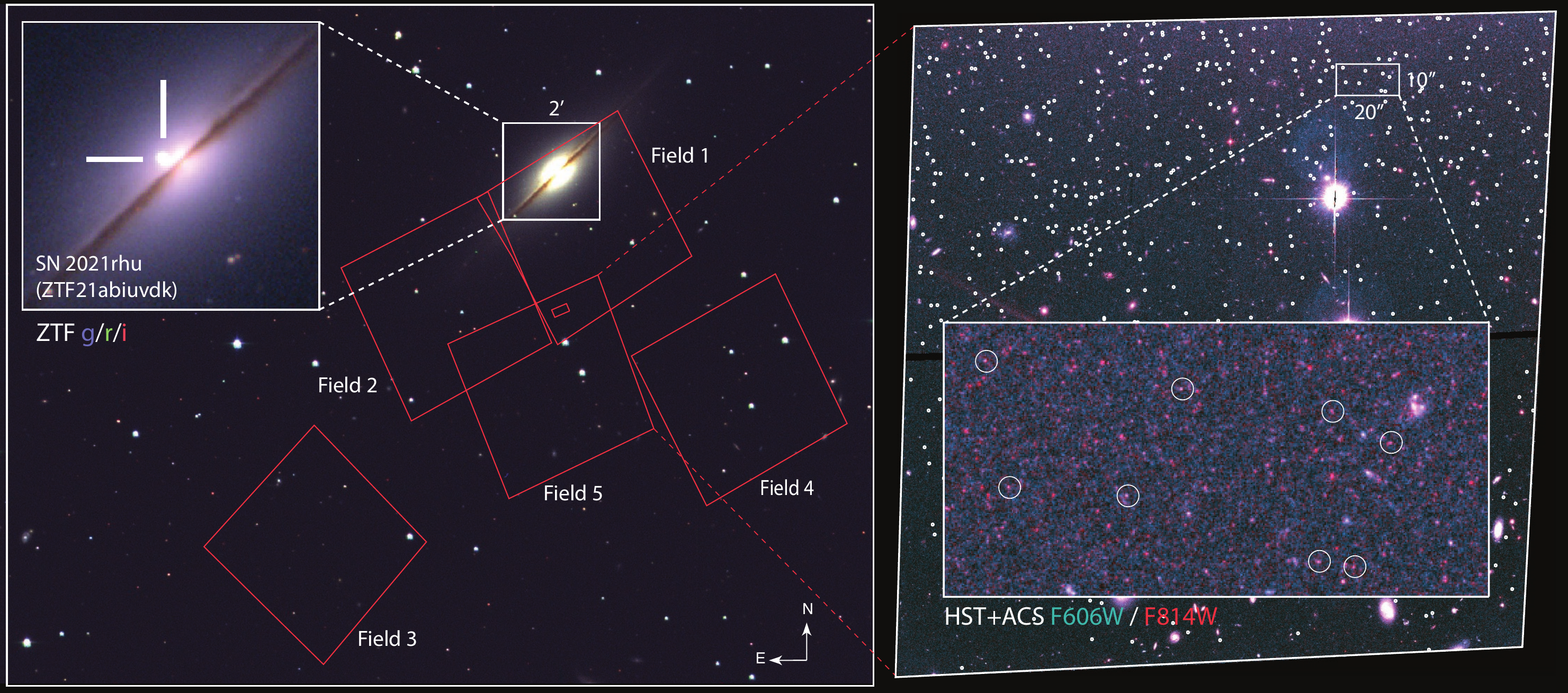}
    \caption{(Left) A combined color image of NGC\,7814 from the ZTF $gri$ data with Fields 1 - 5 from the HST ACS observations overplotted in red. The inset shows the central 2\arcmin$\times$2\arcmin of NGC\,7814 and the   position of SN\,2021rhu (upper left). (Right) HST ACS Field 5 (one of the three fields, along with Fields 3 and 4, used for the distance measurement)in the F606W and F814W filters. The inset panel (bottom left) is a  10\arcsec$\times$20\arcsec region showing individual stars near the tip of the red giant branch magnitude (marked as white circles).}
    \label{fig:hst_ztf}
\end{figure*}
Circumventing the two largest known sources of systematic error is key to achieving the percent level precision in the local distance scale and  resolving the Hubble tension. Firstly, Cepheid variables strongly prefer young, star-forming environments. This has been shown to bias the inferred SN~Ia luminosity, and hence $H_0$ \citep{rigault2020}, though the size of this effect is currently debated \citep{2018ApJ...867..108J}. While the current Cepheid distance ladder,  consisting of SNe~Ia within a 40\, Mpc volume in young, star forming hosts \citep{riess2021:h0}, addresses this issue by evaluating $H_0$ from only the Hubble flow SNe~Ia in low stellar mass hosts, it is important to measure $H_0$ using a volume-limited calibrator and Hubble flow sample of SNe~Ia in all types of host galaxies, to quantify the environment dependent biases, given the profound cosmological implications of the Hubble tension.  
TRGB stars, unlike Cepheid variables, are found in both old and young environments, hence they can probe SN~Ia host galaxies of all morphological types in a given volume. The TRGB is a well-understood standard candle, arising from the core helium flash luminosity at the end phase of red giant branch (RGB) evolution for low-mass stars \citep{2019ApJ...882...34F,2021ApJ...906..125J,2021ApJ...919...16F}. Furthermore, TRGB stars, found in the outskirts of galaxies, are less prone than Cepheids to biases from crowding, and are also comparatively less sensitive to reddening systematics, a potential contribution  to the Cepheid $H_0$ measurements \citep[e.g.,][]{Mortsell2021}. 

Secondly, the current sample of SNe~Ia for $H_0$ measurements is derived from several ($> 20$) different  combinations of telescopes, instruments and filters \citep[e.g.][]{Scolnic2021:p+,riess2021:h0}. Although there have been significant efforts to cross-calibrate the heterogeneous systems \citep{Brout2021:p+}, there are irreducible  uncertainties associated with the data where the filters, instruments and even telescopes no longer exist. In light of these outstanding sources of error, it is beneficial to have a volume limited sample of calibrator and Hubble flow SNe~Ia observed with the \emph{same} instrument.

Addressing these issues, here we present a uniform distance ladder, with both calibrator and Hubble flow SNe~Ia observed by the Zwicky Transient Facility \citep[ZTF;][]{bellm2019,2019PASP..131g8001G}, calibrated on the basis of the TRGB method. As both the calibrator and Hubble flow rungs of the distance ladder are observed with the same instrument, we only rely on a relative photometric calibration, which is a significantly simpler task than controlling the absolute calibration of an SN~Ia sample. In this pilot study, we present  ZTF calibrator SNe~Ia within a nearby volume of luminosity distance, $D_L < 20$ Mpc and measure preliminary distances, where possible, for those SNe~Ia using the tip of the red giant branch. In the long term, we need a ZTF calibrator sample of $\sim$ 100 SNe~Ia  to get to $\sim 1\%$ precision  and accuracy on $H_0$ (assuming the current precision in the TRGB absolute magnitude calibration) to resolve the tension. With the James Webb Space Telescope (JWST) scheduled to start taking data in mid-2022, we can feasibly extend the calibrator rung to $D_L \sim 80\,$Mpc.  ZTF has already observed well-sampled light curves for more than one hundred SNe~Ia in this distance range. Therefore, within the $D_L \leq 80\,$Mpc volume we will no longer be limited by the rate of SNe~Ia in galaxies to obtain calibrator distances, currently  a limiting factor for the largest calibrator sample \citep{riess2021:h0}.

\section{Data and methodology}
\label{sec:method}
We present the data for SNe~Ia observed by ZTF in a $D_L < 20$ Mpc volume, for which there are sufficient observations to infer a distance from the TRGB method. While 5 SNe~Ia have adequate light curve sampling to get precise peak magnitudes, shape and color parameters from SNe~Ia, only one of them, ZTF21abiuvdk (aka SN~2021rhu) has observations of the host galaxy to get an accurate distance.  

SN~2021rhu exploded in NGC~7814 (see Figure~\ref{fig:hst_ztf}, left inset), at $\alpha=0.8143^{\circ}, \delta=16.1457^{\circ}$ (J2000 coordinates), classified as a SN~Ia on the Transient Name Server \citep[TNS;][]{tns_21rhu, SNIascore21}. We obtained photometry with a 1-day cadence for SN~2021rhu with ZTF, in the $g,r,i$ filters between $-14.1$ and +172.5 days. These observations begin on 2021-07-01.4 UTC. Hence, we obtained a densely sampled light curve with the ZTF observing system \citep{2020PASP..132c8001D}, in multiple filters, i.e., the same  system as the Hubble flow sample \citep[as presented in][]{2021arXiv211007256D}. The images were processed with the pipeline as detailed in \citet{2019PASP..131a8003M}. The lightcurve,  thus far, spans a large phase range from 2021-07-01.4 to 2021-11-11.11 (Figure~\ref{fig:saltfit_spectra} shows the range used in the fit). We have also obtained a well-sampled spectral time series, beginning with a classification spectrum with the SEDmachine \citep{2018PASP..130c5003B,2019A&A...627A.115R,2022PASP..134b4505K} on 2021-07-05. These are presented in detail in a companion paper (Harvey et al. in prep). Figure~\ref{fig:saltfit_spectra} shows a maximum-light spectrum obtained with the SPectrograph for the Rapid Acquisition of Transients \citep[SPRAT;][]{sprat2014} on the Liverpool Telescope \citep[LT;][]{lt2004}.


SNe~Ia distances are inferred from light curve peak luminosity, shape and color. The most widely used light-curve fitting algorithm, which we adopt for our analysis, is the Spectral Adaptive Lightcurve Template - 2 \citep[SALT2;][]{guy2007}. This model treats the color entirely empirically, without distinguishing the intrinsic and extrinsic components.  
We use the most updated, published version of SALT2 \citep[SALT2.4, see][]{guy2010,Betoule2014} as implemented in \texttt{sncosmo} 
v2.1.0\footnote{\url{https://sncosmo.readthedocs.io/en/v2.1.x/}} \citep{2016ascl.soft11017B}, identical to the lightcurve inference of the Hubble flow sample in \cite{2021arXiv211007256D}. In the fitting procedure, we correct the SN fluxes for extinction due to dust in the Milky Way (MW). Extinction values for the SN coordinates  derived in \citet{2011ApJ...737..103S} were applied, using the galactic reddening law proposed in \cite{1989ApJ...345..245C}, with a total-to-selective absorption ratio, $R_V = 3.1$, the canonical MW value.
\begin{figure}
    \centering
    \includegraphics[width=.48\textwidth, trim = 0 10 0 10]{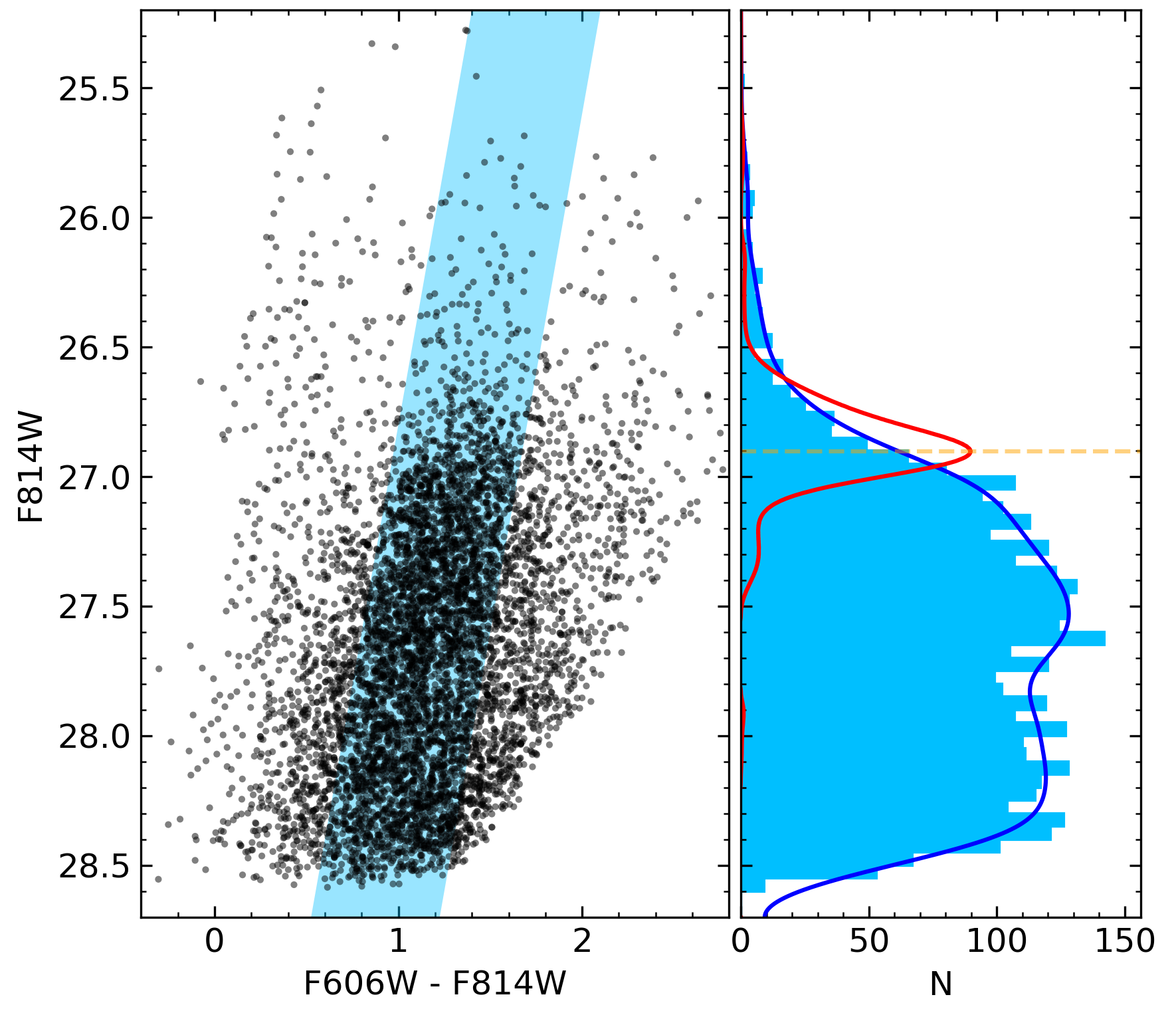} 
    \caption{(Left) The color-magnitude diagram (CMD) for the RGB stars (black points) using the observed magnitudes in the F606W and F814W filters. 
    The blue shaded area represents the color selection for the  metal-poor RGB stars used in the TRGB determination. We show the CMD for Field~4 as an example of one of the fields which are used for the distance estimate in this work. The tip is detected using an edge detection method as detailed in \citet{2021ApJ...906..125J}. (Right) The luminosity function (LF; blue histogram) for the RGB stars with the Gaussian smoothing using a 0.1 mag scale overplotted. The red curve shows the tip detection.}
    \label{fig:trgb_cmd}
\end{figure}
\subsection{TRGB distance estimate} 
NGC~7814 was observed with the Advanced Camera for Surveys (ACS) on HST (see Figure~\ref{fig:hst_ztf}) covering a total of seven fields as part of the GHOSTS survey \citep{radburn-smith2011}. Here, we reanalyse the data using a pipeline by the Carnegie-Chicago Hubble Program \citep[CCHP;][]{2019ApJ...882...34F}   which implements its own point-spread function (PSF) fitting photometry based on DOLPHOT \citep{Dolphin2000} modeling synthetic PSFs with TinyTim \citep{TinyTim2011}.
\begin{table}
    \centering
    \caption{ The statistical and systematic error budget for computing the TRGB, shown for the case of the individual Field~4 as used in our distance determination. 
    }
    \begin{tabular}{|c|c|}
    \hline 
    Effect &  $\sigma$ (mag)\\
    \hline 
    
      Edge Detection &  0.04  \\
      
        Photometry choice   & 0.02
\\
      Color selection    & 0.01
\\
    Smoothing selection     & 0.01
\\
   Empirical Aperture Correction   & 0.01
\\
       ACS Zero-Point    & 0.02
\\
       ACS EE Correction   & 0.02
\\
      \hline
    \end{tabular}
    \label{tab:error_trgb}
\end{table}

The details of the pipeline can be found in \citet{2021ApJ...906..125J}.
We select fields 3, 4,  and 5 from the entire dataset since fields 1 and 2 are close to the disk of the galaxy and hence susceptible to high crowding and extinction biases, whereas fields 6 and 7 are very sparse making it difficult to identify the TRGB. 
We perform artificial star tests, as a robustness test of the photometric pipeline, by injecting  $\sim$200,000 stars into the FLC images and recover them using DOLPHOT. 
The artificial stars have a similar colors range to blue RGB stars  in the shaded region of the CMD (see Figure~2).
We populate stars within a brightness range of $25 < F814W \leq 29$~mag. To mimic the observed spatial distribution and luminosity function (LF), we place more stars in the inner region of the galaxy. 

The LF was binned with a width of 0.01 mag and smoothed with a Gaussian kernel of 0.1 mag (e.g. Figure~\ref{fig:trgb_cmd}). The edge detection is derived from the first derivative of the scale smoothed LF \citep[see][for details]{Hatt2017}.  We perform a tip detection on each individual field and find the values to be consistent within errors and hence,  take the average of the individual measurements with a conservative error for the TRGB.
We find a Milky Way extinction corrected tip at 
$F814W_{0,\rm TRGB} = 26.81 \pm 0.06$ mag.\footnote{Data associated with Figure~\ref{fig:trgb_cmd} can be found in \url{https://github.com/hanlbomi/NGC7814-Field4}}
Details of the tests, the impact of assumptions on the various components of the pipeline and consistency between the individual fields and with distances reported in the literature are presented in a companion paper (Jang et al. 2022 in prep).  
Inferring the distance to NGC 7814 from this tip measurement requires an absolute calibration of the TRGB magnitude. We use the most recent absolute calibration of the TRGB magnitude that is derived from multiple primary anchors, namely MW, LMC. SMC and NGC\,4258, from \citet{2021ApJ...919...16F}, 
 $M_{\rm F814W}^{\rm TRGB} = -4.049\pm0.038$ \citep[see also,][for a new calibration from the Milky Way]{Li2022:TRGB} and 
we obtain a distance modulus of $\mu = 30.86 \pm 0.07$ mag.  The individual sources of error in the final distance uncertainty are presented in Table~\ref{tab:error_trgb}. We note that there is debate in the literature regarding the value of this absolute magnitude, depending on the assumptions in the primary anchors. The TRGB magnitude can be calibrated using the parallax distances from the Gaia satellite early data release 3 \citep{GaiaEDR3:2021} to Milky Way (MW) globular clusters, e.g. $\omega$ Cen, detached eclipsing binary distances to the Large and Small Magellanic Clouds \citep[LMC, SMC][]{2019Natur.567..200P, Gracyk2020} and the water maser distance to the nearby galaxy NGC 4258 \citep{Reid2019}. The difference assumptions / anchor combinations have led to a difference of up to $\sim 0.1$ mag in the inferred absolute magnitude. A summary of the various absolute calibrations in the literature are provided in \citet{Blakeslee2021,2021ApJ...919...16F}.


We combine the calibrator data with the ZTF DR1 Hubble flow sample \citep{2021arXiv211007256D}. TRGB stars are found in all types of SN~Ia host galaxies and therefore, the TRGB-calibrated sample will be volume limited. To have a completely volume-limited distance ladder, i.e. both calibrator and Hubble flow rungs, in this study, we also only fit the volume-limited Hubble flow sample from ZTF DR1.  However, we note that the ZTF Hubble flow SN~Ia sample has been built with carefully controlled selection function via the Bright Transient Survey \citep[e.g.,][]{Fremling2020, 2020ApJ...904...35P}. In future studies, this will be a unique advantage since we can use this to compute the selection effects for SNe~Ia with redshifts up to $z \sim 0.1$, where the impact of peculiar velocity errors is significantly lower than for literature samples \citep[see][for details]{2021arXiv211007256D}. For this study, since we are dominated in the uncertainty budget by having only a single calibrator, we conservatively take the sample to be complete to $z \leq 0.06$. This selection cut reduces the Hubble flow sample from 200 to 98 SNe~Ia.
\begin{figure*}
    \centering
        
    \includegraphics[width=.41\textwidth]{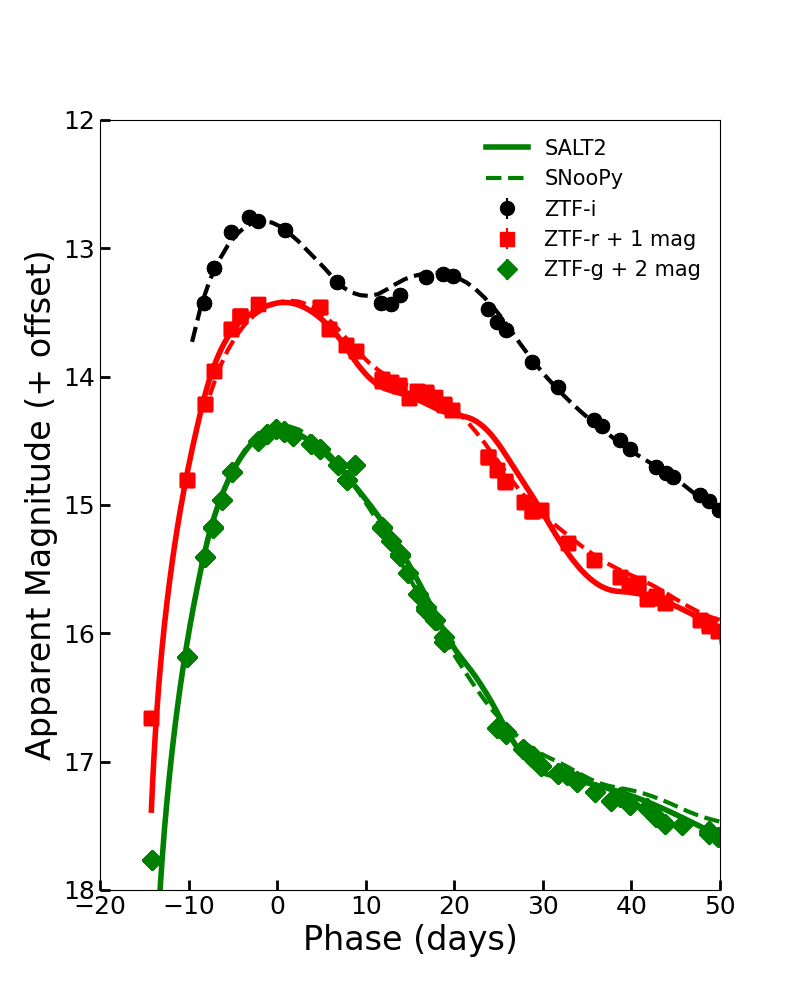}
        \includegraphics[width=0.43\textwidth]{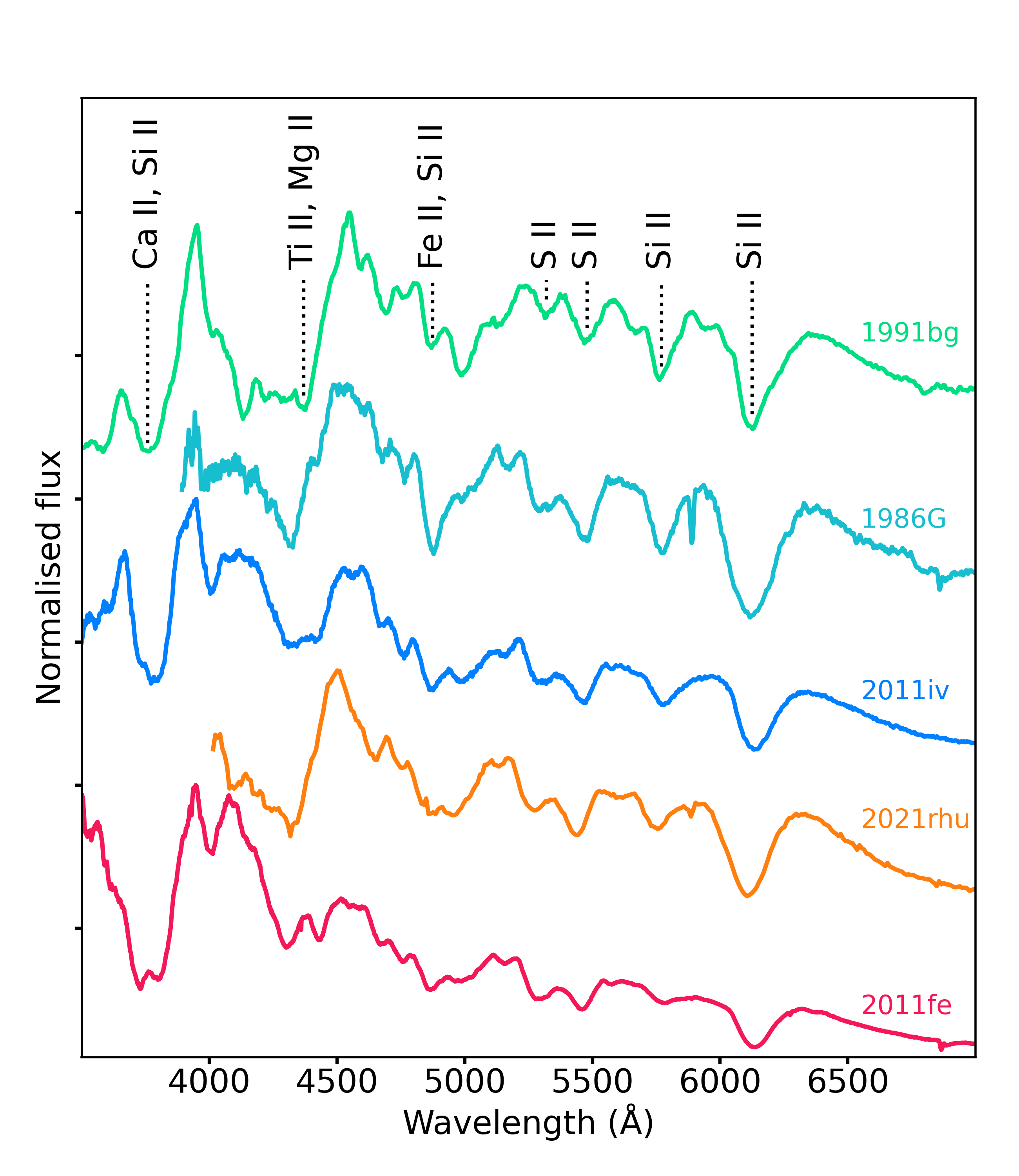}
    \caption{(Left): Lightcurve of SN\,2021rhu in the $g$ (green diamonds),$r$ (red squares) ,$i$ (black circles) filters along with the SALT2 model fit to the $g$,$r$ filters overplotted (solid) and the SNooPy model fit to the $g,r,i$ filters (dashed). The plot has truncated this the phase at which the SALT2 model is defined. (Right) A maximum light spectrum of SN\,2021rhu (orange), in comparison with the peculiar, subluminous SN~1991bg \citep[green;][]{1992AJ....104.1543F}, transitional SN 1986G \citep[cyan;][]{Cristiani1992} and SN~2011iv \citep[blue;][]{Foley2012}, the latter has been used as a calibrator object and the normal SN~2011fe \citep[red;][]{Parrent2012}. The most common spectral features of intermediate mass and iron group elements of SNe~Ia at maximum light are shown as dotted lines.  We find that the near maximum light spectrum of SN~2021rhu is very similar to transitional SNe~Ia (see also Harvey et al. in prep.)}
    \label{fig:saltfit_spectra}
\end{figure*}



\section{Results}
  We fit the SALT2 light-curve model to the calibrator SN  and get the peak luminosity, light-curve width and color. We note that since SALT2 is not well defined at wavelengths redder than 7000 {\AA}, we only fit the $g$ and $r$ filters \citep[e.g.][]{2019ApJ...881...19J}. 
SN 2021rhu has SALT2 parameters $m_B = 12.22 \pm 0.033$, light-curve shape $x_1 = -2.074  \pm 0.025$, and color, $c = 0.054  \pm 0.028$. While the $x_1$ and $c$ are within the range of typical cosmological cuts (e.g., $|x_1| < 3$, $|c| < 0.3$), it has a low $x_1$ value which is also seen in peculiar, fast-declining SNe~Ia. However, the light curves of SN~2021rhu show a clear shoulder in the $r$ band and a second peak in the $i$ band (Figure~\ref{fig:saltfit_spectra}), characteristic of normal and transitional SNe~Ia used for cosmology \citep{Hsiao2015}.   We also compute the color-stretch parameter, $s_{BV}$, with the \texttt{SNooPY} method, since it is shown to be better at parametrizing the fast declining SNe~Ia \citep{Burns2014}. We find $s_{\rm BV} = 0.72$ consistent with normal/transitional SNe~Ia, appropriate to use for cosmology  \citep{Burns2018}. It is also spectroscopically similar to transitional SNe~Ia like SN~2011iv \citep{Foley2012}, which have been used for estimating $H_0$ \citep{2019ApJ...882...34F}, thus this object is consistent with the cosmological sample of SNe~Ia. 

Here, we present the formalism for inferring $H_0$. The absolute magnitude of SNe~Ia, $M_B$, is given by
\begin{equation}
    m_B^0 - \mu_{\rm host} = M_B 
    \label{eq:abs_mag}
\end{equation}
where ${m}_B^0$ is the \emph{standardized} apparent peak magnitude of the SN~Ia and $\mu_{\rm host}$ is the distance modulus to the host galaxy based on the TRGB method.
The Hubble flow SNe~Ia measure the intercept of the magnitude-redshift relation, $a_B$.  Ignoring higher order terms, the intercept is given by
\begin{equation}
\resizebox{.91\hsize}{!}{
$a_B = \log cz + \log \left[1+\frac{(1-q_0)z}{2} - \frac{(1 - q_0 - 3
q^2_0 + j_0)z^2}{6}\right] - 0.2 m_B^{0}$}.
\label{eq:intercept}
\end{equation}
 We fix $q_0$, $j_0$, the deceleration and cosmic jerk parameters to the standard values of $-0.55$ and 1 respectively, since the low-$z$ SN~Ia sample alone cannot constrain them. We note that while cosmological studies with SNe~Ia correct the redshifts for the Hubble flow sample accounting for peculiar motion due to local large scale structure, this effect has been shown to be a sub-dominant source of error in measuring $H_0$ \citep{2021arXiv211003487P}, which is especially true here since only a single calibrator dominates the error budget.   Additionally, we note that while this effect can be important when the  calibrator sample is increased, as we have mentioned above, there will also be a simultaneous increase both in the size and the median redshift of the Hubble flow sample.
$m_B^0$ is expressed in terms of the light-curve parameters and corrections as 
\begin{equation}
    m_B^0 = m_B + \alpha x_1 - \beta c - \delta_{\mu - {\rm bias}}
\label{eq:mb_0}
\end{equation}
where $\alpha$ and $\beta$ are the slopes of the width-luminosity and color-luminosity relations, respectively, and $\delta_{\mu - {\rm bias}}$ is the bias correction needed to account for selection effects and other sources of distance bias. Following the formalism of  \citet{Brout2022:Cosmo}, the canonical term for the host galaxy ``mass-step" correction is absorbed in the bias correction $\delta_{\mu - {\rm bias}}$ \citep[see also][]{BS21}. Since both the calibrator described by equation~\ref{eq:abs_mag} and the Hubble flow SNe~Ia described by equation~\ref{eq:intercept} are constructed to be volume-limited, such that they both have the \emph{same} mass-step correction, the $\delta_{\mu - {\rm bias}}$  term  will cancel out.

\begin{figure*}
    \centering
    \includegraphics[width=.48\textwidth]{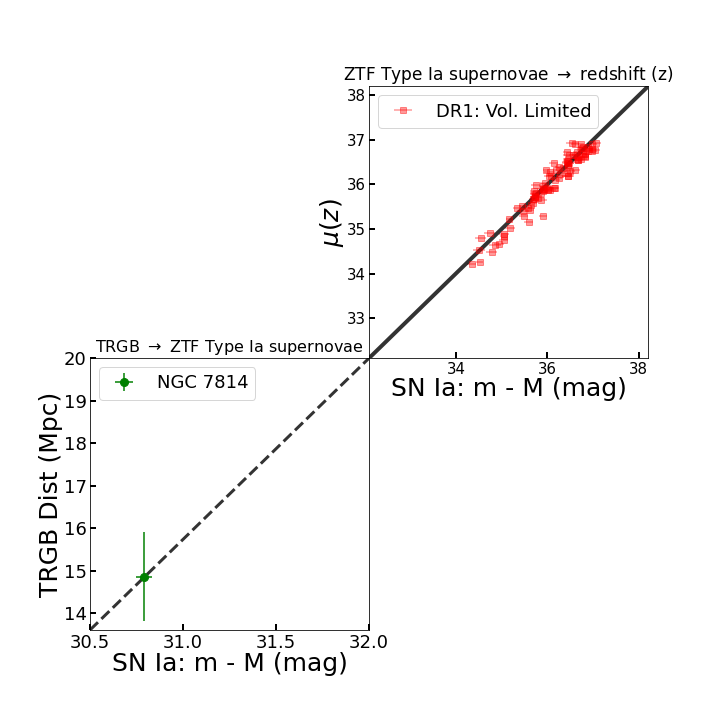}
    \includegraphics[width=.48\textwidth]{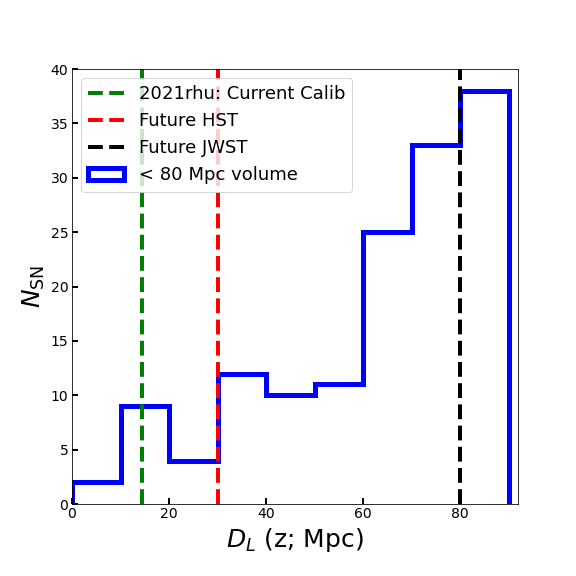}
    \caption{(Left): The current ZTF distance ladder with SN\,2021rhu in NGC 7814 (green; the TRGB distance is plotted in linear scale instead of a distance modulus) and the Volume Limited Hubble flow sample from ZTF DR1 (red). We emphasize that all SNe~Ia in this distance ladder are observed with the same survey. (Right): Histogram of luminosity distances for nearby ($z \leq 0.02$) ZTF SNe~Ia with sufficient observations infer distances. Distances are computed from the redshift assuming standard cosmology from \cite{2020A&A...641A...1P} with $H_0 = 67.4\,$km\,s$^{-1}$\,Mpc$^{-1}$ and $q_0$, $j_0$ of -0.55 and 1 respectively. Hence, they are only indicative.  The distance for the current calibrator and the maximum distance feasible with HST and JWST are plotted as green, red and black vertical dashed lines respectively. There is a total of 114 SNe~Ia with high-quality light curves in this volume, providing a large sample to build a ZTF-only distance ladder.}
    \label{fig:outlook_jwst}
\end{figure*}

The error for each SN includes fit uncertainty from the SALT2 covariance matrix ($\sigma_{\rm fit}$), the peculiar velocity error ($\sigma_{\rm pec}$) and $\sigma_{\rm int}$. 
\begin{equation}    
    \sigma^2_{\rm m} = \sigma^2_{\rm fit} + \sigma^2_{\rm pec} + \sigma^2_{\rm int}
    \label{eq:error_magnitude}
\end{equation}

For $\sigma_{\rm pec}$ we derive the magnitude error from a velocity error of 300 km\, s$^{-1}$ \citep{Carrick2015}.  
We use \texttt{PyMultiNest} \citep{2014A&A...564A.125B}, a python wrapper for \texttt{MultiNest} \citep{2009MNRAS.398.1601F} to derive the posterior distribution on the parameters. With the current calibrator, we find,  $H_0 = 76.94  \pm 6.4\, {\rm km}\,{\rm s^{-1}}\,{\rm Mpc^{-1}}$ . 
We also fit for $H_0$ using the entire Hubble flow DR1 sample and find $H_0 = 77.60 \pm 6.0\, {\rm km}\,{\rm s^{-1}}\,{\rm Mpc^{-1}}$, a small difference of $0.66\,{\rm km}\,{\rm s^{-1}}\,{\rm Mpc^{-1}}$.  
This uncertainty is not significantly smaller when using the entire gold sample for ZTF DR1 compared to the volume limited one. This is because the main source of uncertainty is from having on a single calibrator object.

We also infer the corrected peak magnitudes with \texttt{SNooPy} \citep{Burns2014}. While \texttt{SNooPy} uses a light-curve template, as opposed to a spectral template for SALT2, it is trained with a larger sample of transitional SNe~Ia similar to SN~2021rhu, hence we compare $H_0$ values from both methods. We compute distances to both the Hubble flow SNe~Ia and SN~2021rhu with the \texttt{EBV\_model2}. Using the same analysis method as for the SALT2 fitted distances, we infer an $H_0$ value of  $77.58 \pm 6.1\, {\rm km}\,{\rm s^{-1}}\,{\rm Mpc^{-1}}$, a difference of $0.64 \, {\rm km}\,{\rm s^{-1}}\,{\rm Mpc^{-1}}$ from the value using SALT2. This difference is significantly smaller than the uncertainty on $H_0$ from either method.  Moreover, since  \texttt{SNooPy} has a well-sampled training set to build the $i$-band template, we also infer $H_0$ from the $g,r,i$ filter combination and find a value of $76.17 \pm 6.0\, {\rm km}\,{\rm s^{-1}}\,{\rm Mpc^{-1}}$, a difference of 0.77 ${\rm km}\,{\rm s^{-1}}\,{\rm Mpc^{-1}}$ from the fiducial case.

\begin{table*}
\caption{The contribution from individual terms in the error budget for measure $H_0$ with the current uniform distance ladder and the forecast with expected distances from JWST.}
    \centering
    \begin{tabular}{|c|c|c|}
        \hline
        Quantity &  Current Uncertainty (mag) & Expected Uncertainty (mag) \\
        \hline  
        SN~Ia intrinsic scatter & 0.15 & 0.1 / $\sqrt{100}$ = 0.01 \\ 
     TRGB absolute calibration    & 0.038 & 0.023 \\
     TRGB in SN~Ia hosts & 0.05 & 0.05 / $\sqrt{100}$ = 0.005 \\
    Peculiar Velocity & 0.02 & $0.01$ \\
    Intercept of the Hubble diagram ($5 a_B$) & 0.013 & 0.004 \\
    \hline
    \end{tabular}

    \label{tab:error_ztfjwst}
\end{table*}

\section{Discussion and Conclusion}
\label{sec:discuss}
We present an estimate of $H_0$ from a uniform distance ladder using the same survey for the calibrator sample as a homogeneous, untargeted Hubble flow sample. We use a TRGB distance to a nearby host galaxy of an SN~Ia with high-cadence data in the ZTF $g,r,i$ filters. The current uncertainty is not sufficient to weigh in on the Hubble tension. We note that even a factor of 2 reduction in the Hubble flow sample by imposing the volume limit does not impact the uncertainty on $H_0$; the error currently is driven by having only a single ZTF SNe~Ia with robust, independent distances. However, this can be increased with HST observations for nearby host galaxies. 
In the $D_L < 20$ Mpc volume, one where we can achieve completeness relatively quickly, ZTF has observed 5 more spectroscopically normal SNe~Ia with well-sampled light curves, a sample expected to increase by $\sim 1-2$ per year for the remainder of ZTF operations. These SNe are 
\begin{enumerate}[noitemsep]
    \item ZTF19aacgslb (SN 2019np) in NGC 3254
    \item ZTF20abijfqq (SN~2020nlb) in NGC 4382 (M85)
    \item ZTF20abrjmgi (SN~2020qxp) in NGC 5002
    \item ZTF21aaabvit (SN~2021J) in NGC 4414
    \item ZTF21aaqytjr (SN~2021hiz) in UGC 7513
\end{enumerate}

All the SNe~Ia listed above have coverage in the $g,r,i$ filters beginning from  at least two weeks before maximum light and extending beyond +70 days. We note that even with this small volume sample, there are early-type host galaxies like NGC 4382, for which other methods like Cepheid variables are not viable to obtain distances. 
In this volume, the number of calibrator SNe~Ia is limited by the rate of SNe~Ia exploding in the universe. The ZTF calibrator sample within the 20 Mpc volume, accumulated to-date, is however, sufficient to measure $H_0$ to $\sim 3\%$ accuracy using only HST for TRGB observations.  In our analyses, we only infer the SN~Ia light-curve parameters using $g$ and $r$ filters since SALT2 is not optimal for redder wavebands.  Recently, improved SNe~Ia models, e.g.  SALT3 \citep{Kenworthy2021}, and / or BayeSN \citep{Mandel2020,Thorp2021} have been demonstrated to enable an accurate use of redder wavebands e.g. the ZTF $i$-band. In future work, we will implement these models trained with high-cadence ZTF SN~Ia data in the $g,r,i$ wavebands to measure SN~Ia distances.  It has been demonstrated that the improved SN~Ia models can reduce the SN~Ia intrinsic scatter to $\lesssim 0.1$ mag \citep[e.g., see][]{Mandel2020}. We nominally expect the $\sigma_{\rm int}$ of a similar level when retraining the ZTF calibrator and Hubble flow samples with improved lightcurve models (as assumed for Table~\ref{tab:error_ztfjwst}
 We note that our Hubble flow sample has only $\sim$ 100 SNe~Ia. This is due to two strict cuts. We apply both the restriction on a host spectroscopic redshift and the limit at the volume of $z \leq 0.06$. The entire DR1 sample contains $> 750$ SNe~Ia \citep{2021arXiv211007256D}, which is a significantly larger sample than the current Hubble flow rung in the literature. We have also shown here that $H_0$ inferred from the complete host spectroscopic redshift sample  of SNe~Ia , with a median $z = 0.057$, is consistent with our volume limited subsample. The complete phase-I of ZTF operations has well-sampled lightcurves of $\sim 3000$ of which $\sim 650$ are within the $z \leq 0.06$ volume. These have been discussed in \citet{2021arXiv211007256D} and will be presented in future work as part of the ZTF second data release (DR2; Smith et al. in prep). As discussed above, for the current work the Hubble flow sample is not the limiting factor in the final uncertainty on $H_0$. However, going forwards the Hubble flow sample will be significantly augmented, both in terms of numbers and the maximum redshift to mitigate the impact peculiar velocity uncertainties \citep[e.g., as demonstrated in ][]{2021arXiv211003487P}. We summarise the forecast uncertainties in Table~\ref{tab:error_ztfjwst}.

Future TRGB observations with the near infrared camera (NIRCam) on JWST can extend the calibrator sample volume out to larger distances of up to 80 Mpc. In the volume $20 < D_L < 80$\, Mpc, we have high-cadence light curves of 106 more SNe~Ia  already obtained (see Figure~\ref{fig:outlook_jwst}), expected to increase by the end of ZTF. Therefore, the complete sample of ZTF SNe~Ia in a volume where JWST observations are feasible can increase the current calibrator sample by a factor of $\sim 2-3$. We emphasize that current SN~Ia cosmology requires cross-calibrating several heterogeneous photometric systems \citep{Brout2021:p+}. To get to percent level precision, it is an important cross-check to have observations of a large sample of SNe~Ia on a single photometric system, that is the \emph{same} for  calibrator and Hubble flow SNe~Ia. 

 While other concluded and/or ongoing SN~Ia surveys have also observed SNe~Ia in the distance range feasible for JWST, ZTF has two key advantages. Firstly, the prospective calibrator sample of ZTF is $\sim$ factor two larger in the same volume, since other surveys typically have $\sim 50 - 60$ SNe~Ia in the specific distance range. This is important to increase the statistical power in the calibrator dataset compared to the current calibrator sample. Secondly, there is a sizable Hubble flow sample of $\sim 3000$ SNe~Ia, with a maximum redshift of $z = 0.1$ \citep[as discussed in][and Smith et al. in preparation for the second data release]{2021arXiv211007256D} on the \emph{same} photometric system as the calibrator sample, allowing us to reduce systematic errors both from cross-calibration uncertainties and dependence on the host galaxy environment. 
 
In addition to the improvements in the second and third rungs via the single system SN~Ia sample proposed here, we would also expect improvements in the absolute calibration of the TRGB magnitude from a combination of multiple primary anchors. For the MW, an increase number of observations and a new full-scale astrometric solution in Gaia DR3 (and subsequently, DR4), compared to  EDR3 \citep{Lindegren2021} will decrease the random and systematic errors.  Moreover, potential improvements in the extinction maps and primary distance calibration can also decrease the systematic errors in the calibration of the TRGB absolute magnitude from the small and large Magellanic clouds compared to current measurements \citep{Hoyt2021}. For NGC4258, with the maser distance, the current error budget in \citet{2021ApJ...906..125J} is conservative. Improvements in the zeropoint of the F814W filter and its EE correction as well as new observations of multiple halo fields of NGC4258 \citep{Hoyt2021} can appreciably reduce the systematic error. Assuming a 0.04 mag error in each of the primary calibrations, we expect a calibration of the TRGB magnitude to have a uncertainty of 0.023 mag (Table~\ref{tab:error_ztfjwst})
More anchor galaxies, e.g. Sculptor \citep{Tran2022} and Fornax \citep{Oakes2022}, can further reduce the uncertainty on the TRGB zero point,  especially with 1\% parallaxes in the future  \citep[see][for details]{2021ApJ...919...16F}.    Future facilities can also increase the number of maser galaxies to calibrate the TRGB absolute magnitude by observing ideal candidates at significantly larger distances than the current, single calibrator galaxy, NGC 4258.
 In the case of the improved TRGB absolute magnitude calibration, this uniform distance ladder can measure $H_0$ at the $\sim 1.3 \%$ level. However, the expected error with the current uncertainty on $M_{TRGB}$ would be $\sim 1.8\%$, hence, sufficient for an independent TRGB estimate of $H_0$, and arbitrating the Hubble tension. 
Hence, the increased statistical power and reduced systematic uncertainties from a single, untargeted survey, make this an ideal approach to resolve the $H_0$ tension.

\section*{Acknowledgements}
Based on observations obtained with the Samuel Oschin Telescope 48-inch and the 60-inch Telescope at the Palomar Observatory as part of the Zwicky Transient Facility project. ZTF is supported by the National Science Foundation under Grant No. AST-2034437 and a collaboration including Caltech, IPAC, the Weizmann Institute for Science, the Oskar Klein Center at Stockholm University, the University of Maryland, Deutsches Elektronen-Synchrotron and Humboldt University, the TANGO Consortium of Taiwan, the University of Wisconsin at Milwaukee, Trinity College Dublin, Lawrence Livermore National Laboratories, IN2P3, France, the University of Warwick, the University of Bochum, and Northwestern University. Operations are conducted by COO, IPAC, and UW. 
SEDMachine is based upon work supported by the National Science Foundation under Grant No. 1106171.  SD acknowledges support from the Marie Curie Individual Fellowship under grant ID 890695 and a Junior Research Fellowship at Lucy Cavendish College. AG acknowledges support from the Swedish Research Council under Dnr VR 2020-03444 and the Swedish National Space Board. This project has received funding from the European Research Council (ERC) under the European Union's Horizon 2020 research and innovation program (grant agreement n 759194 - USNAC). Y.-L.K. acknowledges support by the Science and Technology Facilities Council [grant number ST/V000713/1]. WLF acknowledges support from program \#13691  provided by NASA through a grant from the Space Telescope Science Institute, which is operated by the Association of Universities for Research in Astronomy, Inc., under NASA contract NASA 5-26555. M.~W.~C acknowledges support from the National Science Foundation with
grant numbers PHY-2010970 and OAC-2117997.
This work was supported by the GROWTH Marshal project \citep{Kasliwal2019} funded by the National Science Foundation under Grant No 1545949.

\bibliographystyle{aasjournal}
\bibliography{trgb_ztf}

\end{document}